\documentclass[jkps,nofootinbib,twocolumn,showkeys,superscriptaddress]{revtex4-2}
\usepackage{graphicx}
\usepackage{amssymb}
\usepackage{amsmath}
\usepackage{bm}
\usepackage{url}
\usepackage{color} 
\usepackage{kotex}
\usepackage[normalem]{ulem}

\begin{document}
\title[]{Revisiting small-world network models: Exploring technical realizations and the equivalence of the Newman-Watts and Harary models}
\author{Seora \surname{Son}}
\affiliation{Department of Physics, Gyeongsang National University, Jinju 52828, Korea}
\author{Eun Ji \surname{Choi}}
\affiliation{Department of Physics, Gyeongsang National University, Jinju 52828, Korea}
\author{Sang Hoon \surname{Lee}}
\email[Corresponding author: ]{lshlj82@gnu.ac.kr}
\affiliation{Department of Physics, Gyeongsang National University, Jinju 52828, Korea}
\affiliation{Research Institute of Natural Science, Gyeongsang National University, Jinju 52828, Korea}
\affiliation{Future Convergence Technology Research Institute, Gyeongsang National University, Jinju 52849, Korea}

\date{\today}

\begin{abstract}
We address the relatively less known facts on the equivalence and technical realizations surrounding two network models showing the ``small-world'' property, namely the Newman-Watts and the Harary models. We provide the most accurate (in terms of faithfulness to the original literature) versions of these models to clarify the deviation from them existing in their variants adopted in one of the most popular network analysis packages. The difference in technical realizations of those models could be conceived as minor details, but we discover significantly notable changes caused by the possibly inadvertent modification. For the Harary model, the stochasticity in the original formulation allows a much wider range of the clustering coefficient and the average shortest path length. For the Newman-Watts model, due to the drastically different degree distributions, the clustering coefficient can also be affected, which is verified by our higher-order analytic derivation. During the process, we discover the equivalence of the Newman-Watts (better known in the network science or physics community) and the Harary (better known in the graph theory or mathematics community) models under a specific condition of restricted parity in variables, which would bridge the two relatively independently developed models in different fields. Our result highlights the importance of each detailed step in constructing network models and the possibility of deeply related models, even if they might initially appear distinct in terms of the time period or the academic disciplines from which they emerged.
\end{abstract}


\maketitle

\section{\label{sec:intro}Introduction}

A mixture of regular and random elements in network models, particularly the ``small-world'' network model by Watts and Strogatz (WS)~\cite{Watts1998}, has played a significant role in pioneering the modern era in the field of network science~\cite{NewmanBook1stEd}. The concept of the small world is a crucial aspect of network science, as it describes a phenomenon observed in various real-world networks. This concept captures the occurrence of short average path lengths and high clustering coefficients in networks, indicating that nodes are connected by relatively few steps while still maintaining local interconnectedness.

The WS network is a widely recognized model that combines regularity and randomness~\cite{Watts1998}. It is initialized as a regular lattice, usually the one-dimensional (1D) ring with the periodic boundary condition, with $n$ nodes connected to their $k$ (usually an even number for symmetry) nearest neighbors in the Euclidean sense. Then, a fraction of the original edges are randomly rewired (one end node of each edge is fixed, and the new other end node is chosen uniformly at random) with a probability $p$, introducing randomness into the network. This rewiring process creates shortcuts between nodes, drastically reducing the average shortest path length $L$ and transforming the network into a small world where $L \propto \log n$ in sharp contrast to the case of regular lattices ($L \propto n^{1/d}$) in the Euclidean space with a finite dimension $d$.
In subsequent research~\cite{Newman1999}, Newman and Watts (NW) proposed a modified version of the original Watts-Strogatz model to conduct a simpler renormalization analysis by removing the possibility of network fragmentation. Instead of \emph{rewiring} existing edges, they introduced the modified procedure that a certain number of shortcuts (the fraction $p$ of the original number of edges) are \emph{added} uniformly at random. This process ensures that, on average, $nkp/2$ shortcuts are added to the original regular lattice.

36 years before the WS paper~\cite{Watts1998} basically pioneered the ``digital'' era of network science, in 1962, Harary proposed a graph-generating scheme to optimize the connectivity\footnote{The word ``connectivity'' is occasionally used in the networks literature as a synonym for ``degree'' (the number of nearest neighbors of a node) as well---they should be strictly distinguished in this paper dealing with the Harary graph model~\cite{Harary1962}, of course.} of a graph\footnote{We use the terms ``graph'' and ``network'' interchangeably throughout the paper, which is a common practice in this highly interdisciplinary field.} by considering a predetermined set of numbers of nodes and edges~\cite{Harary1962}. Connectivity, an essential network property better known in graph theory and computer science, refers to the minimal number of elements (nodes or edges) whose removal results in a disconnected network~\cite{Whitney1932}. The theory of network flow problems is closely related to this concept, as a graph's connectivity indicates its robustness and transportation efficiency. In this work, for the first time to our knowledge (based on the fact that the WS and NW papers~\cite{Watts1998,Newman1999}, along with most network science papers~\cite{NewmanBook1stEd}, do not cite the Harary graph~\cite{Harary1962}), we reveal that not only does the Harary graph achieve maximum connectivity, but it can also exhibit small-world characteristics; in fact, under an appropriate condition, the Harary graph and the Newman-Watts network become equivalent to each other. 

In addition, through our journey to discover the equivalence between the two models, we have found that there are nontrivial differences in technical realizations between the original models (both Harary and NW models) and actual mechanisms adopted in numerical tools, representatively, arguably the most popular \texttt{python} package for network analysis: \texttt{NetworkX}~\cite{NetworkX}. For the Harary graph, the stochasticity in the original model~\cite{Harary1962} is not realized, and for the NW model, the random edge addition process poses more restriction in choosing random pairs of nodes than the original model~\cite{Newman1999}. Both alterations cause structural differences: obvious ones for the stochastic Harary graph and the drastically different forms of degree distributions for the NW model. 

We would like to report our findings on these small-world network models to the network science community. In particular, the developers of \texttt{NetworkX}~\cite{NetworkX}, who have contributed tremendously in advancing network science, may consider revising their codes (or providing alternate versions for the original models) for these network models in the future, reflecting our suggestions. Our paper is organized for this purpose. In Sec.~\ref{sec:Harary}, we review the Harary graph and disclose the difference between the original model and the numerical realizations. In Sec.~\ref{sec:NewmanWatts}, we review the NW model and highlight the seemingly subtle difference in random edge addition, which eventually causes significant changes in the degree distribution and notable differences in the average clustering coefficient. We present the equivalence relation between the two models in Sec.~\ref{sec:Harary_vs_NW}, and conclude the paper in Sec.~\ref{sec:conclusion}.

\section{\label{sec:Harary}The Harary graph}

The Harary graph was developed to solve a practical optimization problem---the tradeoff between robustness and the cost of creating edges. Building a network robust to failure or attack~\cite{Albert2000,Cohen2000,Cohen2001} is intuitively related to the connectivity (the minimal number of nodes or edges whose removal results in a disconnected network) of a graph by definition~\cite{Berge1958}, denoted by $r$ in this paper. However, achieving high connectivity typically comes at an inevitable additional cost, specifically regarding the number of edges in the network. The problem of finding the maximum connectivity of any graph with a given number of nodes and edges was posed by Berge~\cite{Berge1958} in 1958 and solved by Harary~\cite{Harary1962} in 1962. In particular, the deterministic version of the Harary graph is known for interesting properties in regard to connectivity, e.g., its connectivity is equal to its minimum degree, which is expressed as ``maximally connected.''~\cite{Gross2004}. 

Unfortunately, the Harary graph is relatively less known in the field of network science, especially in the physics-based community. The \texttt{NetworkX} package~\cite{NetworkX}, though, implements the Harary graph. Still, perhaps because its deterministic version is much more popular than its stochastic version in the original paper due to the former's interesting properties mentioned in the previous paragraph~\cite{Harary1962}, the package misses a crucial aspect of the Harary graph: the possibility of stochastic elements in it. Restricting the Harary graph only to the deterministic regime hinders properly showcasing its diverse structures from stochasticity. To unleash the issue in this paper, we distinguish two different realizations of the Harary model, namely the ``original Harary (OH)'' model and the ``\texttt{NetworkX} Harary (XH)'' model. The OH model refers to the model as originally conceived by Harary in his seminal paper, while the XH model refers to the version implemented in \texttt{NetworkX}~\cite{NetworkX}. 

An intriguing feature of the OH model is its ability to generate both deterministic and probabilistic graphs, depending on the given number of edges. Contrary to the common misconception (shared by the latest version 3.1 of \texttt{NetworkX}~\cite{NetworkX} at the time of writing) that the Harary model only yields deterministic graphs, it possesses the flexibility to generate graphs that incorporate probabilistic elements based on the given edges according to the original model description~\cite{Harary1962}. We hope that our work clarifies the situation in the context of modern network science, particularly by inspecting its close relation to small-world-type network models~\cite{Watts1998,Newman1999} introduced near the turn of this century, later in the paper (Sec.~\ref{sec:Harary_vs_NW}). 

\subsection{\label{sec:originalHarary}The original Harary graph: The possibility of stochasticity}

\begin{figure*}
\begin{tabular}{lll}
(a) & (b) & (c) \\
\includegraphics[width=0.3\textwidth]{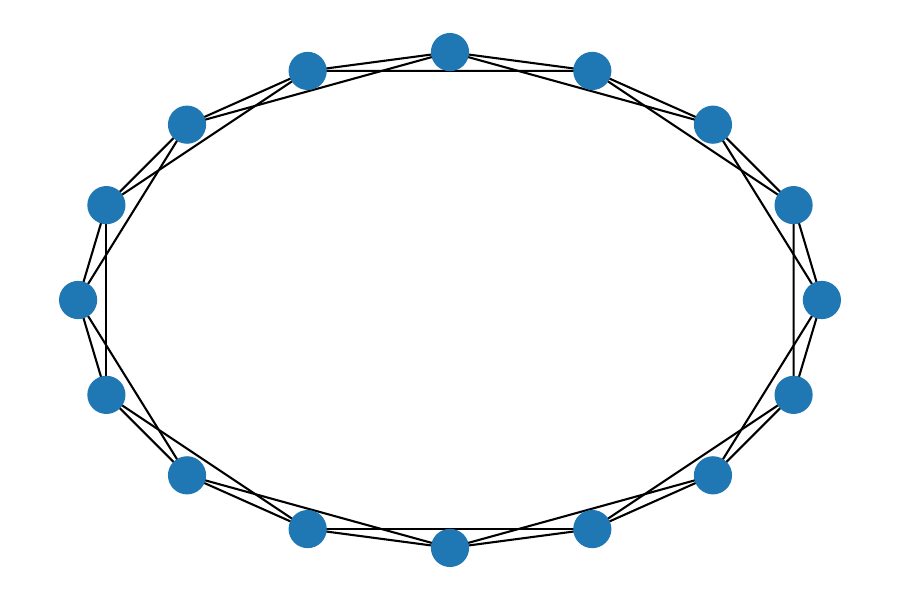} & 
\includegraphics[width=0.3\textwidth]{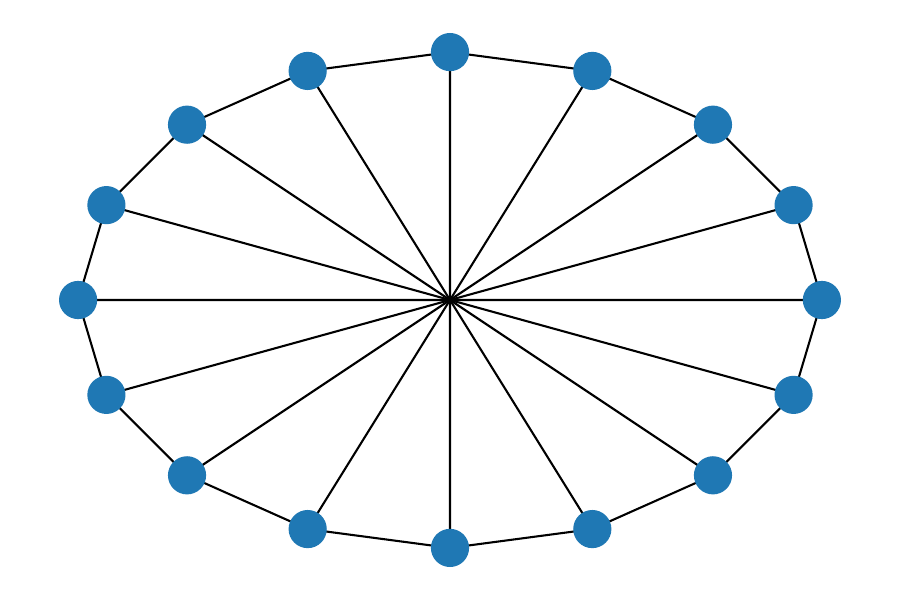} &
\includegraphics[width=0.3\textwidth]{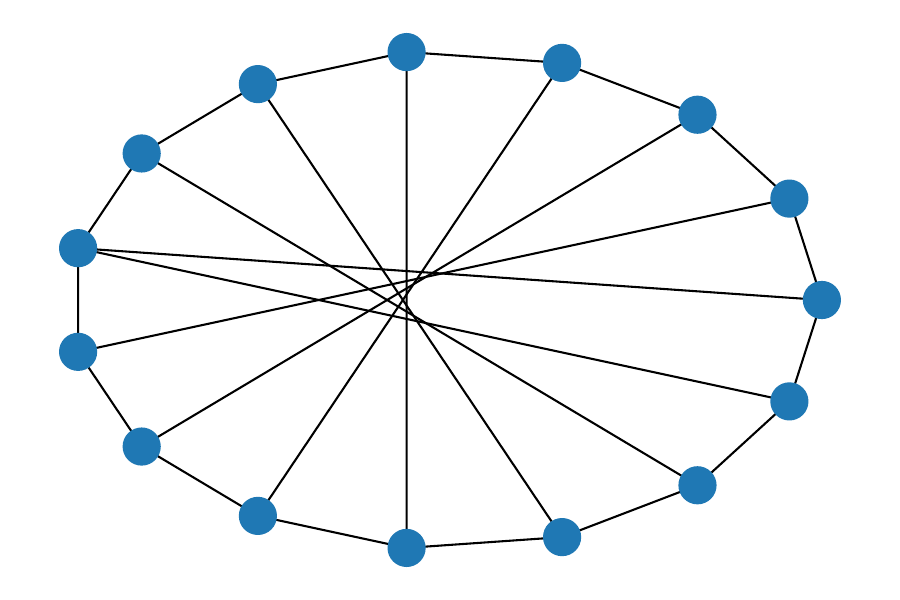} \\
(d) & (e) & (f) \\
\includegraphics[width=0.3\textwidth]{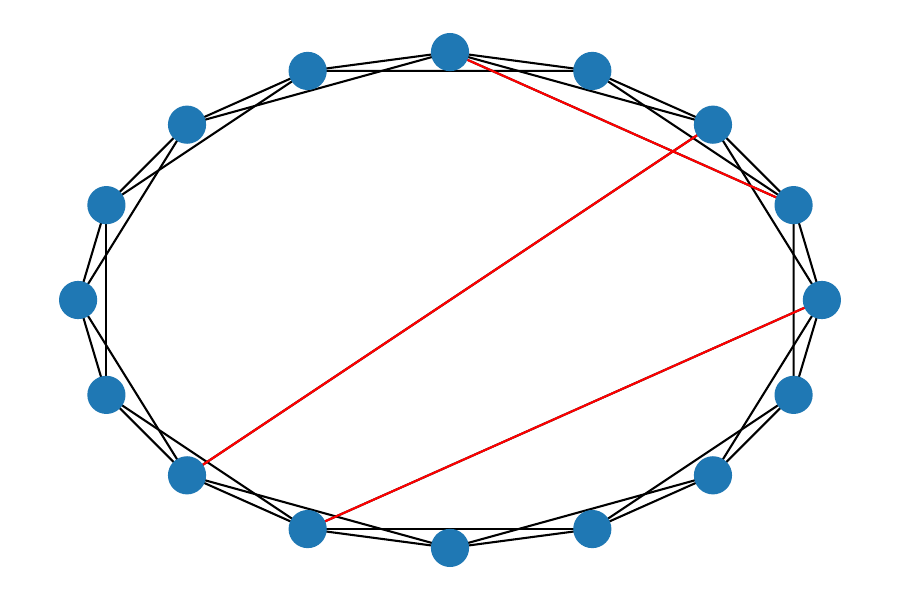} &
\includegraphics[width=0.3\textwidth]{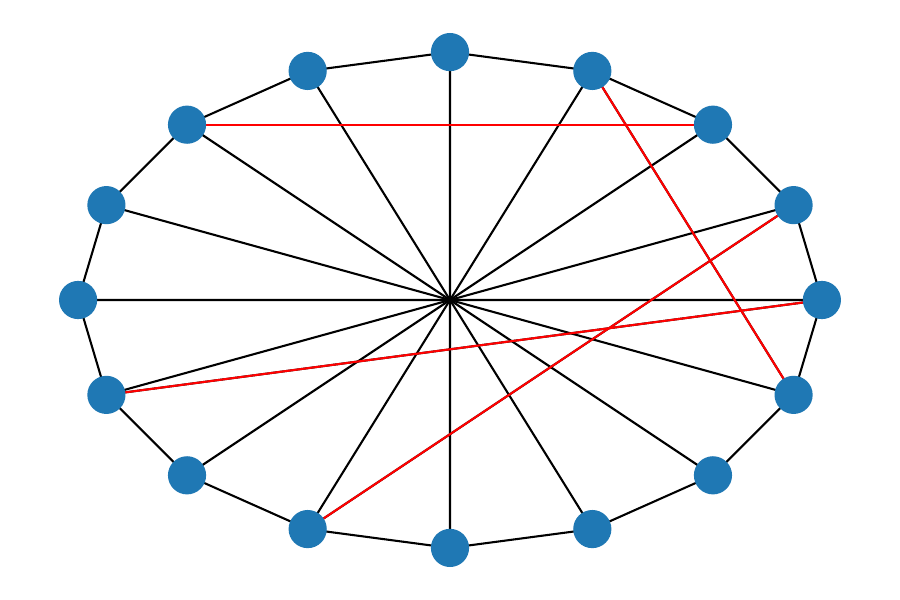} &
\includegraphics[width=0.3\textwidth]{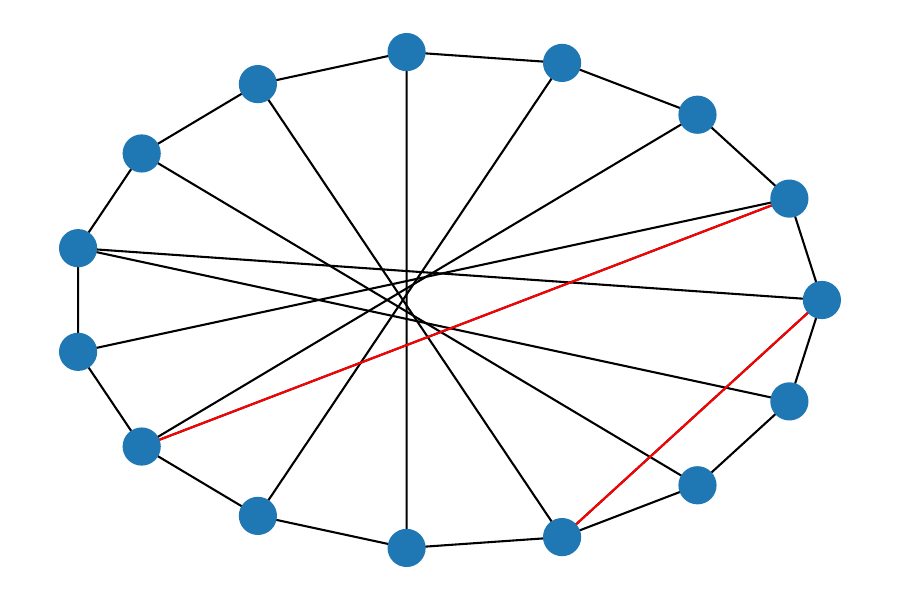}
\end{tabular}
\caption{Examples of the original Harary (OH) graph listed as the cases in Sec.~\ref{sec:originalHarary}. See the descriptions at the end of each case for the parameter values. Note that each graph in the lower panels (d), (e), and (f) has extra random edges or ``shortcuts'' (the red lines) added to the deterministic graph (the black lines) shown in the upper panels (a), (b), and (c).}
\label{fig:Harary_examples}
\end{figure*}

The OH model introduced in Ref.~\cite{Harary1962} aims at generating a network with the maximum connectivity $r$ for given numbers $n$ and $m$ of nodes and edges, respectively. First, it is proved that
\begin{equation}
r = \left\lfloor \frac{2m}{n} \right\rfloor \,,
\label{eq:r_condition}
\end{equation}
where $\lfloor x \rfloor$ is the floor function (the greatest integer less than or equal to $x$)~\cite{Harary1962}. The OH model generates deterministic graphs when the remainder of $2m/n$, denoted by $2m\%n$, is $0$ or $1$, and stochastic graphs otherwise. We list the step-by-step construction procedure for all of the possible cases depending on different combinations of $n$ and $m$, in the following. Figure~\ref{fig:Harary_examples} presents examples of all of the cases.
\begin{enumerate}
\item When $2m/n$ is an integer, i.e., $r = 2m/n$:
\label{case:2m_n_integer}
\begin{enumerate}
\item When $r$ is even: it involves creating a regular ring graph with $n$ nodes and $rn/2$ edges, i.e., connecting all of the node pairs $(i,j)$ with the distance $d$ where $1 \le d \le r/2$, which is essentially the same as creating a WS (NW) model~\cite{Watts1998,Newman1999} without any probability for the edge rewiring (addition) process, respectively. The resulting graph, denoted by $H_{2d}$, is regular with the unique degree $k = r \equiv 2d$ and $m=nd$ edges. Figure~\ref{fig:Harary_examples}(a) shows an example with $n=16$, $m=32$, and $r=4$.
\label{case:2m_n_integer_r_even}
\item When $r \equiv 2d+1$ is odd: the graph $H_{2d}$ mentioned in the case~(\ref{case:2m_n_integer_r_even}) is constructed first, followed by adding $n/2$ edges connecting the diametrically opposite nodes, i.e., the node pairs $(i,j)$ with $|i-j| = n/2$, in the initial $H_{2d}$ to complete $H_{2d+1}$, which is regular with the unique degree $k = r = 2d+1$. Note that the diametrically opposite nodes are always uniquely determined because the number $n$ of nodes should be even ($\because 2m = nr$ and $r$ is odd). Figure~\ref{fig:Harary_examples}(b) shows an example with $n=16$, $m=24$, and $r=3$.
\label{case:2m_n_integer_r_odd}
\end{enumerate}

\item When $2m\%n = 1$, i.e., $2m/n = r + 1/n$, then $nr$ must be odd ($\because 2m = nr + 1$):
\label{case:nr_odd_2m_n_1}
\begin{enumerate}
\item A regular graph, $H_{r-1}$ (with $r-1$ being even), is taken with $n$ nodes. Then, $(n+1)/2$ new edges are added by connecting pairs of nodes $i$ and $j$ such that $|i-j| = (n-1)/2$. The resulting graph, $H_{r}$, has $(n-1)$ nodes with degree $r$ and a single node with degree $(r+1)$. In fact, the single special node can be any node, so one may call the graph stochastic if nodes are distinguishable from each other, but it is \emph{topologically deterministic} if we do not care about the identity of each node, which is usually the attitude we take in graph theory. Figure~\ref{fig:Harary_examples}(c) shows an example with $n=15$, $m=23$, and $r=3$.
\label{case:2m_n_integer_r_odd}
\label{case:nr_odd_2m_n_1_only_subcase}
\end{enumerate}

\item When $2m\%n > 1$, i.e., $2m/n = r + q/n$ where $q = 2, \dots, n-1$:
\label{case:2m_n_non_integer_2m_n_gtr_1}
\begin{enumerate}
\item When $nr$ is even, i.e., $q$ is even: the construction method from the case~(\ref{case:2m_n_integer}) is applied, either with $n$ or $r$ being even. A regular graph $H_{r}$ with $n$ nodes and $rn/2$ edges is deterministically constructed, and $m - rn/2 = q/2$ remaining edges are added \emph{uniformly at random} to complete $m$ edges in total. It is worth noting that in this subcase, where $r$ is even in particular, the construction mechanism is identical to that of the original NW model~\cite{Newman1999}. Figure~\ref{fig:Harary_examples}(d) shows an example with $n=16$, $m=35$, $r=4$, and $q=6$, which leads to $q/2=3$ remaining edges. Figure~\ref{fig:Harary_examples}(e) shows another example with $n=16$, $m=28$, $r=3$, and $q=8$, which leads to $8/2=4$ remaining edges. 
\label{case:2m_n_non_integer_2m_n_gtr_1_nr_even}
\item When $nr$ is odd, i.e., $q$ is odd: the construction method from the case~(\ref{case:nr_odd_2m_n_1}) is used to create a deterministic graph, $H_{r}$, and $m-(rn+1)/2 = (q-1)/2$ remaining edges are added \emph{uniformly at random} to complete $m$ edges in total. Figure~\ref{fig:Harary_examples}(f) shows an example with $n=15$, $m=25$, $r=3$, and $q=5$, which leads to $(q-1)/2=2$ remaining edges. 
\label{case:2m_n_non_integer_2m_n_gtr_1_nr_odd}
\end{enumerate}
\end{enumerate}

\begin{figure*}
\begin{tabular}{ll}
(a) & (b) \\
\includegraphics[width=0.5\textwidth]{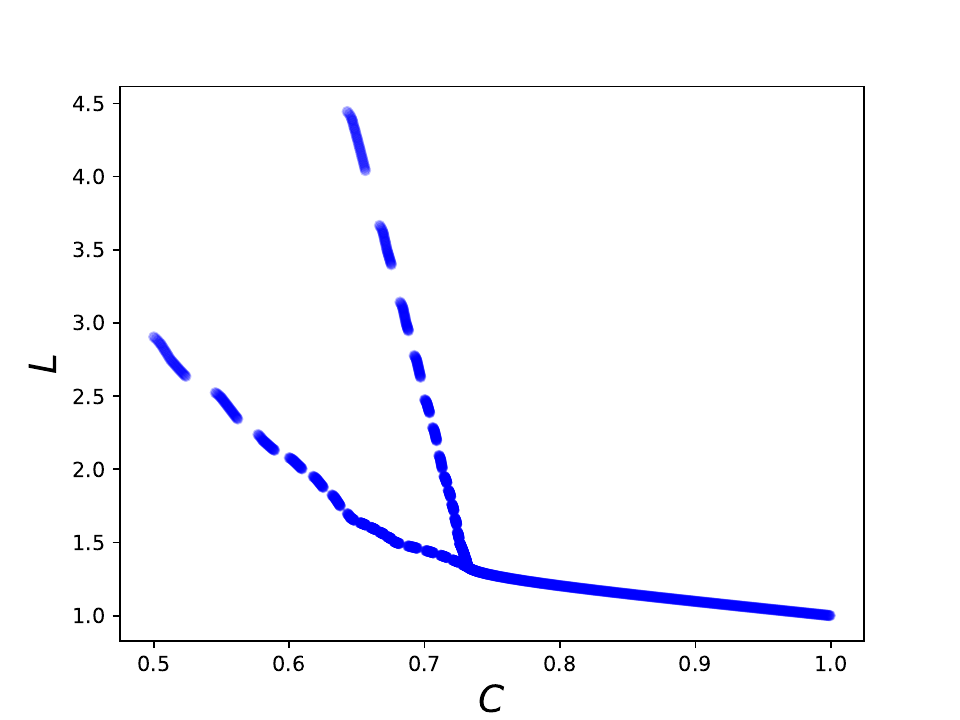} &
\includegraphics[width=0.5\textwidth]{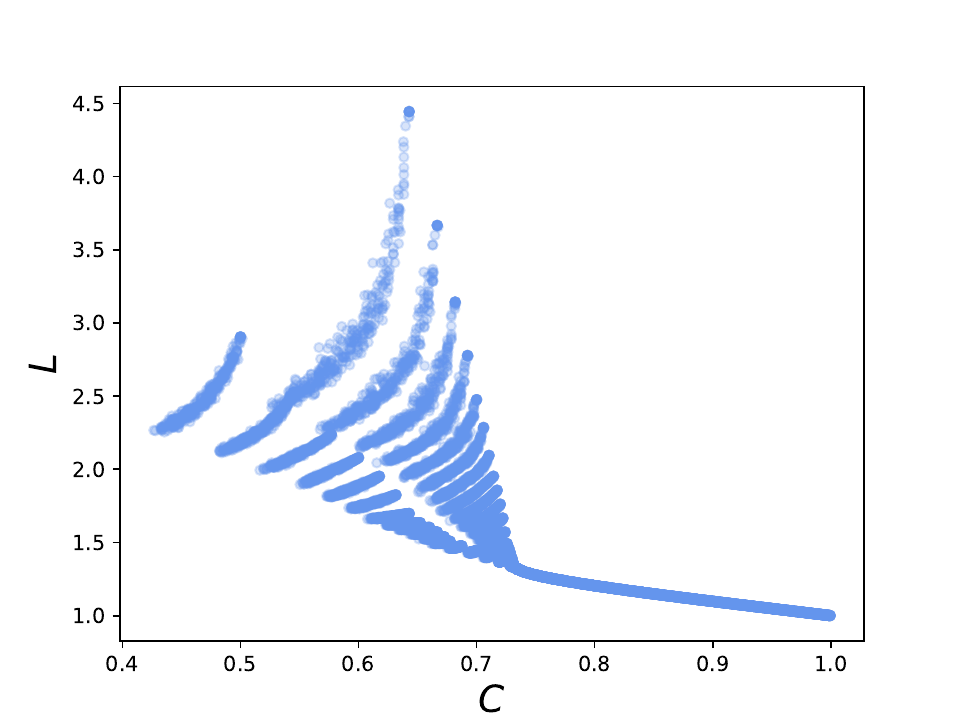}
\end{tabular}
\caption{Graphs generated by two different Harary graph generators: (a) the XH model using the \texttt{hnm\_harary\_graph} function in \texttt{NetworkX}~\cite{NetworkX} and (b) the original OH model with stochastic elements~\cite{Harary1962}, located in the measure space composed of the average clustering coefficient $C$ and average shortest path length $L$. }
\label{fig:Harary_LC}
\end{figure*}

In a nutshell, the Harary graph adheres to a simple yet effective rule: construct the graph deterministically until the maximum connectivity $r = \lfloor 2m / n \rfloor$ is reached based on the given numbers of nodes and edges, and any additional edges are located completely randomly. This rule enables the Harary model to generate both deterministic and probabilistic graphs, depending on the number $n$ of nodes and the number $m$ of edges specified.

\subsection{\label{sec:NetworkXHarary}The NetworkX Harary graph: The completely deterministic graph}

In the \texttt{NetworkX} package, the Harary graph generator offers two types: \texttt{hkn\_harary\_graph} and \texttt{hnm\_harary\_graph}. The former is defined based on a specified value of connectivity $r$ (= \texttt{k} in the \texttt{NetworkX} code)\footnote{Note that the \texttt{NetworkX} code takes the convention of using the letter \texttt{k} for connectivity, for which we use the symbol $r$ throughout our paper to avoid the confusion with the degree.} and the number $n$ of nodes (minimizing the number $m$ of edges), while the latter is determined by the number $n$ of nodes and the number $m$ of edges (maximizing the connectivity $r$). One may assume that the \texttt{hnm\_harary\_graph} (denoted by the XH model from now on) generates the original Harary graph~\cite{Harary1962}, but the part where the randomness is involved, namely the case~(\ref{case:2m_n_non_integer_2m_n_gtr_1}) in Sec.~\ref{sec:originalHarary}, is \emph{replaced with a completely deterministic addition} of new edges. 

We illustrate the difference between the XH and OH models in Fig.~\ref{fig:Harary_LC} by plotting the values of the average clustering coefficient $C$ and the average shortest path length $L$, which are the most prominent properties in the small-world networks~\cite{Watts1998,graph2nn}. 
Figure~\ref{fig:Harary_LC}(a) illustrates the limited range of deterministic XH model, as only a single deterministic graph can be generated for given numbers of $n$ and $m$. On the other hand, when the randomness of the OH model is properly incorporated, as shown in Fig.~\ref{fig:Harary_LC}(b), it becomes possible to generate an ensemble of different graphs taking a wider space of $(C,L)$ for each set of parameters. For a given number $n = 64$ of nodes, we generate $1760$ deterministic XH graphs corresponding to all of the possible integer values of $m \in [256,2016)$, corresponding to $r \in [8,63)$, and $17\,600$ stochastic OH graphs for the same parameter range ($10$ independent realizations for each parameter $m$). 

In other words, we observe that the OH model significantly expands the graph measure space with respect to the XH model by stochasticity, thereby offering broader possibilities for application, such as neural network (NN) model search for machine learning (ML)~\cite{graph2nn}. In fact, we intentionally use exactly the same parameters for the Harary graph used in Ref.~\cite{graph2nn}, which explores the possibility of different graph structures for more efficient NN-based ML. Comparing Fig.~\ref{fig:Harary_LC} in this paper to Fig.~3 of Ref.~\cite{graph2nn}, we confirm that the authors of Ref.~\cite{graph2nn} used the XH model indeed\footnote{The authors of Ref.~\cite{graph2nn} published their codes at \url{https://github.com/facebookresearch/graph2nn}, so we also confirm that they actually used the \texttt{hnm\_harary\_graph} function in \texttt{NetworkX}.}, and we believe that they would have explored a much wider range of the $(C,L)$ space even with the Harary graph if they had adopted the OH model, e.g., Fig.~\ref{fig:Harary_LC}(b).

\section{\label{sec:NewmanWatts}The Newman-Watts model}

Similar to the difference in realizations of the Harary graph in Sec.~\ref{sec:Harary}, we also find the two different generating mechanisms in the NW model~\cite{Newman1999}. We refer to them as the ``original Newman-Watts (ONW)'' model and the ``\texttt{NetworkX} Newman-Watts (XNW)'' model. The term ONW denotes the model as initially proposed by Newman and Watts, while the XNW model represents a version realized by the \texttt{NetworkX} developers\footnote{The developers even left a comment: ``\texttt{\# is that the correct NWS model?}'' in the source code of the \texttt{newman$\_$watts$\_$strogatz$\_$graph} function (\texttt{NetworkX} version 3.1, released on April 4, 2023). Through this paper, we clarify the situation.}. From now on, we present the result of a seemingly small difference in the generating procedure at first glance, which causes quite a notable structural difference.

Note that while the ONW model, originally proposed by Newman and Watts~\cite{Newman1999}, allows for multiple edges and self-loops. However, in this paper, we choose to disallow them for a fair comparison of edge addition mechanisms between the two versions, as the XNW model is designed to construct simple graphs.

\subsection{\label{sec:originalNW}The original Newman-Watts model: A completely random edge addition}

The ONW model is the union of a circulant graph (1D ring) and the classical Erd\H{o}s-R{\'e}nyi (ER) type random graph~\cite{Erdos1959} (the version of Gilbert~\cite{Gilbert1959} with the connection probability, instead of the fixed number of edges in the ER model~\cite{Erdos1959})~\cite{Gu2012}. Initially, the circulant graph is composed of $n$ nodes and $m$ edges, by connecting $k_\mathrm{init} = 2m_\mathrm{init}/n$ nearest nodes from each node. Then, for each given edge in the ring graph, a new edge is added with a probability of $p$ between a randomly chosen pair of nodes. Therefore, the average number of additional edges created is $pm_\mathrm{init}$, which results in the average number $m = (1+p)m_\mathrm{init}$ of total edges. 

Note that the part ``for each given edge in the ring graph'' is completely unnecessary in fact. What actually happens is just adding $pm$ edges uniformly at random on average, and even the case of $p > 1$ is possible unless we strictly impose the interpretation of $p$ as a probability variable. Therefore, the original WS model~\cite{Watts1998} and the ONW model~\cite{Newman1999} have more different aspects than one might initially conceive, which can be crucial when we discuss the \texttt{NetworkX} version in Sec.~\ref{sec:NetworkXNW}. For instance, in contrast to the WS model~\cite{Watts1998}, the ONW model can even show the non-monotonic behavior of the average clustering coefficient as $p$ increases (as the edge density becomes increased, in contrast to the WS model without any change in the edge density), which will be presented in Secs.~\ref{sec:DegDist} and \ref{sec:NW_non_monotonic} with our more accurate analytic derivation than some previous attempts by other researchers.

\subsection{\label{sec:NetworkXNW}The NetworkX Newman-Watts model: A more restricted version}

According to the documentation of the function \texttt{newman$\_$watts$\_$strogatz$\_$graph} in \texttt{NetworkX} (XNW), the process of creating shortcuts involves adding new edges in the following manner: For every edge $(u, v)$ in the underlying $n$-ring with $k_\mathrm{init}$ nearest neighbors, there is a probability $p$ of adding a new edge $(u, w)$ with a randomly selected existing node $w$. Note that in contrast to the ONW model~\cite{Newman1999} in Sec.~\ref{sec:originalNW}, each of the new edges must start from one end node $u$ of an edge $(u,v)$ in the $n$-ring. If we compare this rule with the ONW model without imposing such a restriction, this specific addition rule of XNW induces a more strict version of the edge-addition trial: for every node, there is \emph{a fixed number of trials} ($= k_\mathrm{init}/2$, as each node is chosen at exactly half of the attached edges in the $n$-ring according to the \texttt{NetworkX} code) for edge addition; in contrast, for the ONW model, there is a finite chance for \emph{a node is never selected for the edge-addition process even at} $p = 1$, for example. One can expect that the discrepancy between the two models would become prominent as $p \to 1$, which will be shown indeed in Sec.~\ref{sec:DegDist}.

In retrospect, compared with the WS model~\cite{Watts1998} (and its \texttt{NetworkX} implementation \texttt{watts\_strogatz\_graph}), it seems that the authors of the code \texttt{newman$\_$watts$\_$strogatz$\_$graph} have made minimal modifications to the WS model by replacing the rewiring process of the WS model [for the edge $(u,v)$ in the $n$-ring, creating a new edge $(u,w)$ \emph{with} the removal of the edge $(u,v)$] with the edge-addition process [for the edge $(u,v)$ in the $n$-ring, creating a new edge $(u,w)$ \emph{without} the removal of the edge $(u,v)$]. Note that in contrast to the ONW model described in Sec.~\ref{sec:originalNW}, the control parameter $p$ in the XNW model should be strictly used as a probability variable, as the potential starting point of edge addition is one of the node attached to each edge visited exactly once throughout a realization, which causes nontrivial structural differences from the original NW model that will be presented from now on. 

\subsection{\label{sec:DegDist}The degree distribution and its effect on clustering coefficient}

\begin{figure*}
\begin{tabular}{ll}
(a) & (b) \\
\includegraphics[width=0.5\textwidth]{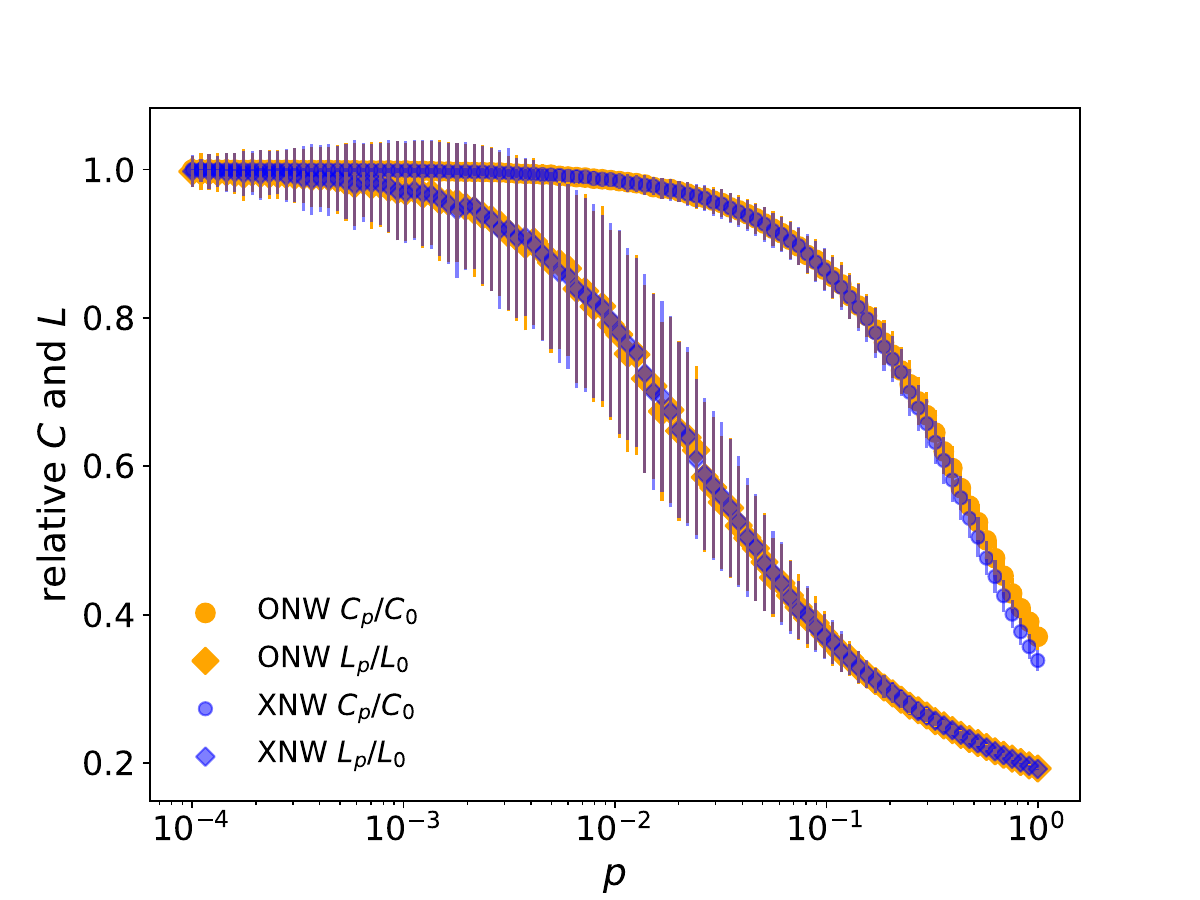} &
\includegraphics[width=0.5\textwidth]{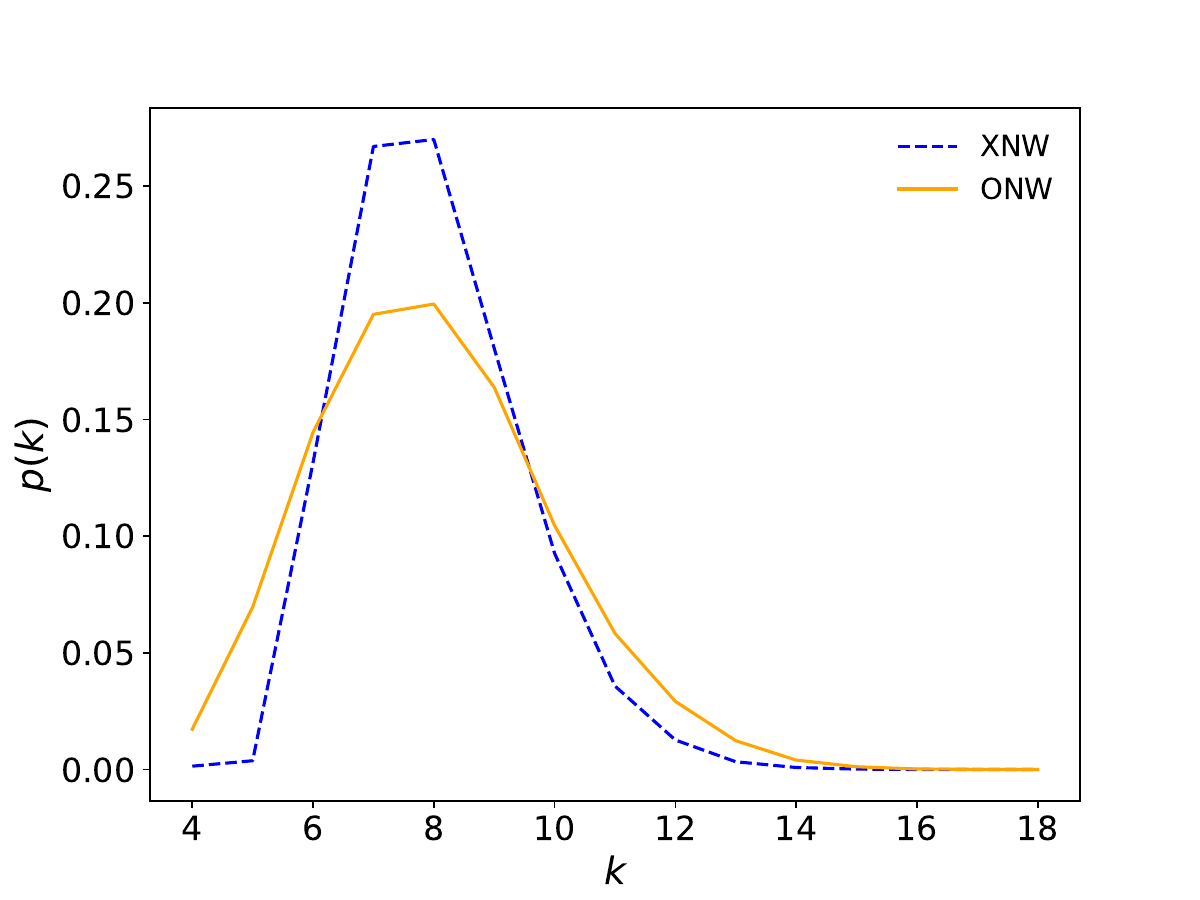} \\
(c) & (d) \\
\includegraphics[width=0.5\textwidth]{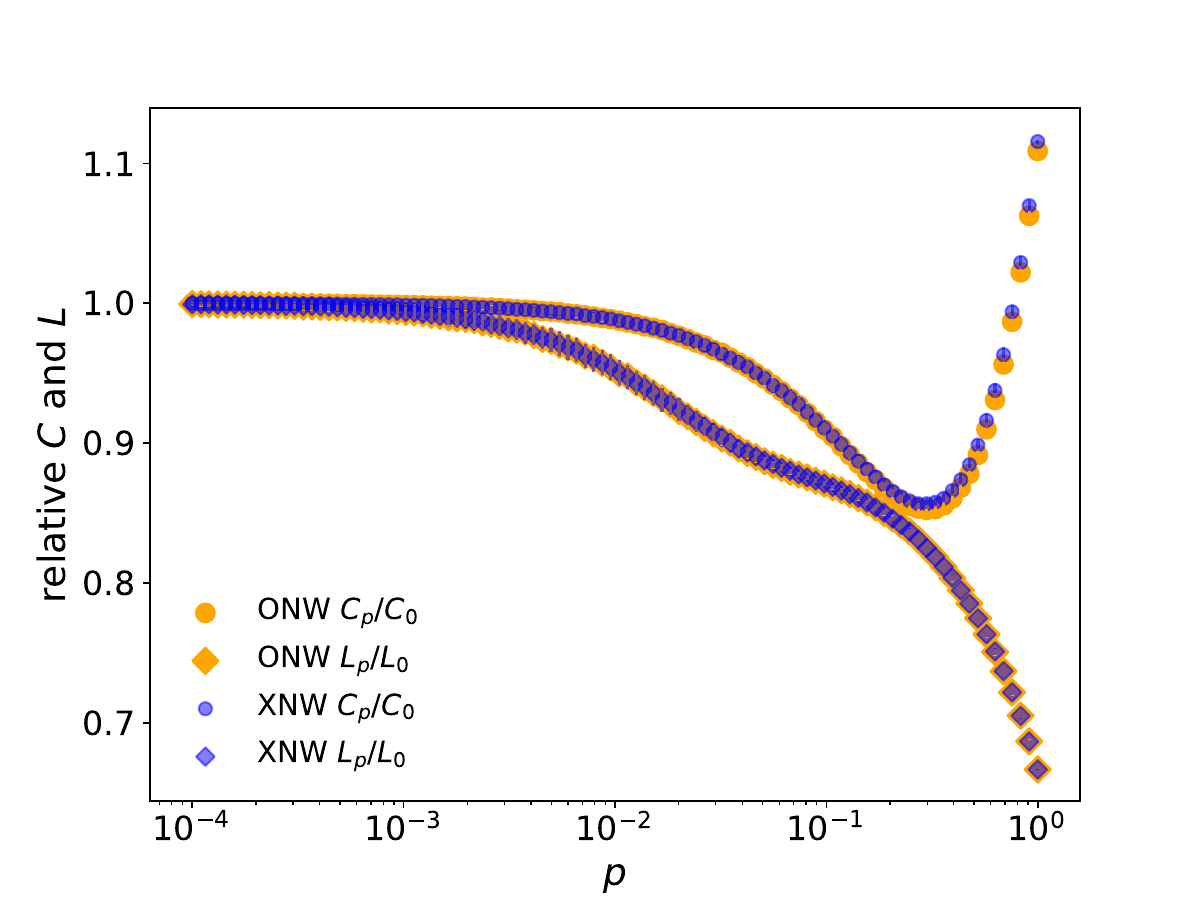} &
\includegraphics[width=0.5\textwidth]{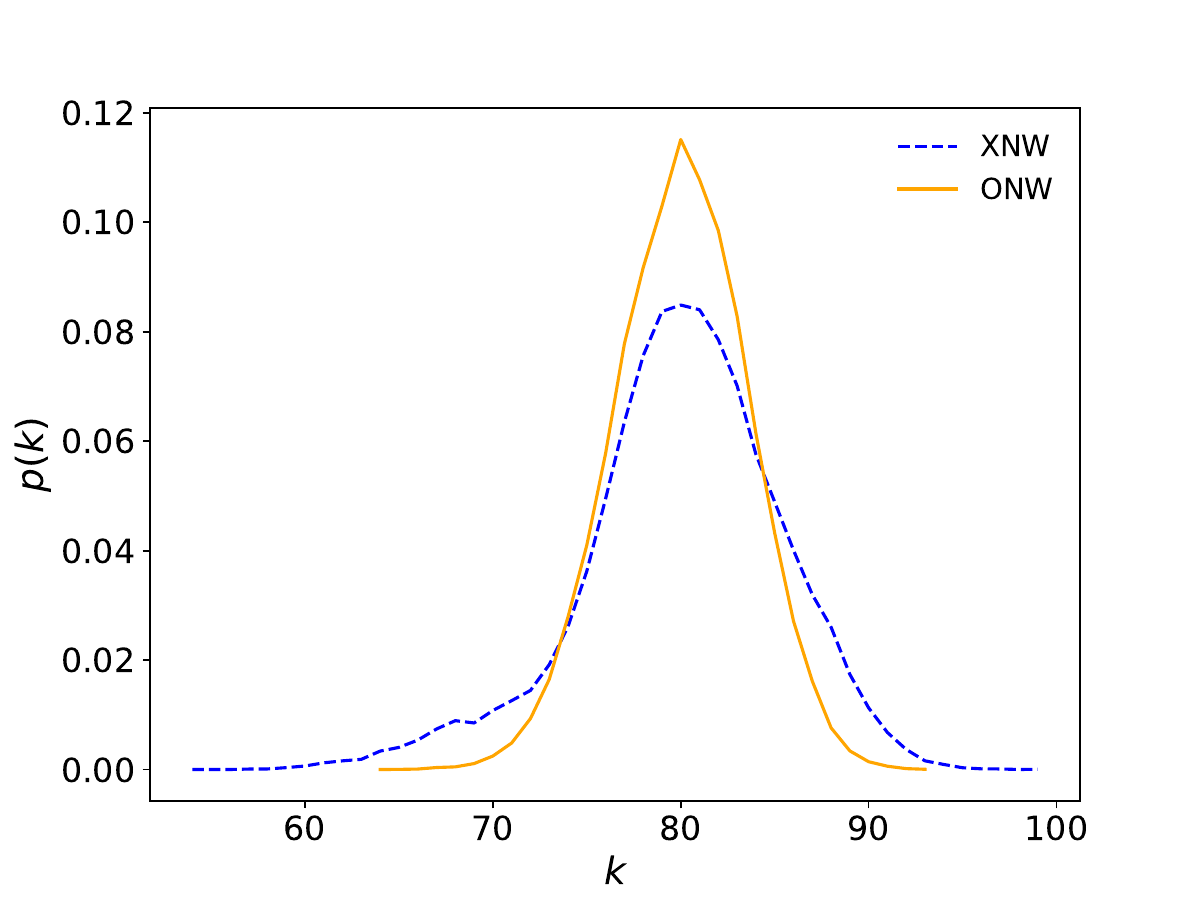}
\end{tabular}
\caption{(a) The relative clustering coefficient $C_p/C_0$ and the relative average shortest path length $L_p/L_0$, with respect to the 1D ring case $p=0$, of the ONW and XNW models with $n = 100$ nodes at $k_\mathrm{init}=4$ for different values of the addition probability $p$, (b) the degree distribution of the ONW and XNW models with $n = 100$ nodes at $k_\mathrm{init}=4$ and $p=1$, (c) the same plot as panel (a) with $n = 100$ and $k_\mathrm{init}=40$, and (d) the same plot as panel (b) with $n = 100$ and $k_\mathrm{init}=40$. All of the plots are from $500$ independent realizations of each parameter value for a given network; the error bars in panels (a) and (c) represent the standard deviation over the realizations and the degree distributions in panels (b) and (d) show the normalized frequencies of each degree value for the entire set of the realizations.}
\label{fig:CC_L_pk}
\end{figure*}

\begin{table}
\caption{The standard deviation of the degree distribution for the ONW and XNW models with different edge densities, corresponding to the cases ($n = 100$ and $p = 1$) in Figs.~\ref{fig:CC_L_pk}(b) and \ref{fig:CC_L_pk}(d).} 
\label{table:stdev}
\begin{ruledtabular}
\begin{tabular}{lcc}
models & $k_\mathrm{init} = 4$ ($\langle k \rangle = 8$) & $k_\mathrm{init} = 40$ ($\langle k \rangle = 80$) \\
\hline
ONW & $4.32$ & $8.66$ \\
XNW  & $3.74$ & $13.03$ \\
\end{tabular}
\end{ruledtabular}
\end{table}

The simplest but fundamental structural difference between the ONW and XNW models is the degree distribution, considering the difference in choosing nodes for edge addition described in Secs.~\ref{sec:originalNW} and \ref{sec:NetworkXNW}. The degree distribution affects various other structural properties of networks, including one of the most important characteristics of this type of small-world network: the average clustering coefficient (the fraction of connected pairs between the neighbors of each node, averaged over all of the nodes)~\cite{NewmanBook1stEd}. We observe subtle differences in the average clustering coefficient $C$ between the ONW and XNW models when the probability $p \to 1$, depending on the initial degree $k$, as shown in Fig.~\ref{fig:CC_L_pk} (the case of $n = 100$). When $k_\mathrm{init} = 4$ and $p \to 1$, $C$ of the ONW model is higher than that of the XNW model in Fig.~\ref{fig:CC_L_pk}(a), while when $k_\mathrm{init} = 40$, $C$ of the ONW model is lower than that of the XNW model in Fig.~\ref{fig:CC_L_pk}(c). In contrast, the average shortest path length is unaffected by the difference in generative mechanisms of the ONW and XNW models, all the way up to $p=1$.

We believe that the difference in the average clustering coefficient is related to the different degree distributions for the ONW and XNW models. In order to systematically investigate this, we examine the degree distributions of the ONW and XNW models when $p = 1$. The degree distributions of both models have the same average value $\langle k \rangle = 2 k_\mathrm{init}$ as $p = 1$, but the distribution of the ONW model is broader than that of the XNW model  when $k_\mathrm{init} = 4$, as shown in Fig.~\ref{fig:CC_L_pk}(b) and Table~\ref{table:stdev}. On the other hand, the degree distribution of the XNW model is broader than that of the ONW model when $k_\mathrm{init} = 40$, as shown in  Fig.~\ref{fig:CC_L_pk}(d) and Table~\ref{table:stdev}. 

To inspect the situation, let us focus on the modified part of the degree distribution (MPDD) by the randomly added edges, denoted by $p_{\mathrm{added}}(k)$ from now on. The ONW and XNW models are similar but generate networks using distinct mechanisms. The ONW model adds edges uniformly at random with probability $p$ to a regular graph, so on average, $p k_\mathrm{init}$ extra degrees are added on each node. In contrast, we can treat the XNW model's MPDD as the following two-step process. First, exactly $k_\mathrm{init} /2$ ``stubs'' are added to each node with the probability $p$, and then the $p k_\mathrm{init} /2$ stubs on average are connected to randomly chosen nodes, which completes distributing $p k_\mathrm{init} /2 + p k_\mathrm{init} /2 = p k_\mathrm{init}$ extra degrees to each node.

Let us write down the difference described above more quantitatively, in the case $p = 1$ for simplicity. The MPDD in the ONW model follows the binomial distribution~\cite{Erdos1959,Gilbert1959} with the connection probability $k_\mathrm{init} / (n - 1 - k_\mathrm{init})$ (= the edge density of the initial $n$-ring), namely, 
\begin{equation}
p_{\mathrm{added}}^{\mathrm{ONW}}(k_\mathrm{extra}) = B \left( n,\frac{k_\mathrm{init}}{n-1-k_\mathrm{init}} \right) \,.
\label{eq:original_binomial}
\end{equation}
However, due to the aforementioned constraint (at least $k/2$ edges should be additional attached to each node by default and the rest of $k/2$ edges are distributed randomly in the ER or Gilbert fashion), the MPDD of the XNW model is given by
\begin{equation}
p_{\mathrm{added}}^{\mathrm{XNW}}(k_\mathrm{extra}) = \frac{k}{2} + B\left(n,\frac{k_\mathrm{init}}{2(n-1-k_\mathrm{init})}\right) \,.
\label{eq:NetworkX_binomial}
\end{equation} 
Therefore, according to the characteristics of the binomial distribution, the standard deviations of the MPDD in the two versions of the NW model are different. For the ONW model, 
\begin{equation}
\sigma_{\mathrm{added}}^{\mathrm{ONW}} = \sqrt{\frac{nk_\mathrm{init}}{n-1-k_\mathrm{init}}\left( 1 - \frac{k_\mathrm{init}}{n-1-k_\mathrm{init}}  \right)} \,,
\label{eq:original_pk_stdev}
\end{equation}
while for the XNW model,
\begin{equation}
\sigma_{\mathrm{added}}^{\mathrm{XNW}} = \sqrt{\frac{nk_\mathrm{init}}{2(n-1-k_\mathrm{init})}\left( 1 - \frac{k_\mathrm{init}}{2(n-1-k_\mathrm{init})} \right)} \,.
\label{eq:NetworkX_pk_stdev}
\end{equation}

Defining $s \equiv k_\mathrm{init}/(n-1-k_\mathrm{init})$ for notational convenience, the standard deviations given by Eqs.~\eqref{eq:original_pk_stdev} and \eqref{eq:NetworkX_pk_stdev} become identical at $s = 2/3$. Around this special value $s = 2/3$, for $s < 2/3$ ($s > 2/3$), the standard deviation of the ONW model is larger (smaller) than that of the XNW model, respectively. Therefore, at relatively small $k_\mathrm{init}$, the ONW model exhibits a broader distribution than the XNW model as shown in Fig.~\ref{fig:CC_L_pk}(b), while at relatively large $k_\mathrm{init}$, the XNW model demonstrates a broader distribution than the ONW as shown in Fig.~\ref{fig:CC_L_pk}(d). An intuitively comprehensible example is the absence (presence) of the node with $k = k_\mathrm{init} = 4$ for the XNW (ONW) model, respectively, as depicted in Fig.~\ref{fig:CC_L_pk}(b). As described in Sec.~\ref{sec:NetworkXNW}, the nodes with the original degree $k_\mathrm{init}$ (not selected at all for edge addition) are impossible for $p=1$ in the case of XNW, while those nodes can exist in the case of ONW.

Then, why is the average clustering coefficient larger for the ONW than the XNW for a small $k_\mathrm{init}$ value and the other way around for a large $k_\mathrm{init}$ value? We suspect that a broader degree distribution implies nodes with less altered local clustering structures, highlighted in the case of $k = k_\mathrm{init} = 4$ from the previous paragraph. Of course, the nodes with larger degrees are also more abundant in a broader degree distribution, but their effect in reducing the average clustering coefficient seems to be less prominent. The net result of those effects is not that drastic as illustrated in Figs.~\ref{fig:CC_L_pk}(b) and \ref{fig:CC_L_pk}(d), but they obviously exist (their difference is clearly out of the error bar ranges). 

Finally, we would like to address a noticeable phenomenon happening only in the NW model~\cite{Newman1999}, which is absent in the WS model~\cite{Watts1998}. As you can see in Fig.~\ref{fig:CC_L_pk}(d), from a particular threshold value of the edge addition probability $p$, the average clustering coefficient $C$ starts to \emph{increase} even when $p$ increases. This non-monotonic behavior of $C$ is impossible for the WS model~\cite{Watts1998}, as the smallest value of $C$ occurs at the maximum rewiring probability $=1$ while the number of edges is conserved. In contrast, as the number of edges keeps increasing for increasing values of $p$ in the NW model, it is certainly possible to have the increasing value of $C$, if the rate of emergence of triangles is high enough to defeat the rate of emergence of uncompleted triads. This phenomenon, despite its intuitive theoretical possibility and experimental facts, has never been reported in the context of the NW model to our knowledge, so we explicitly present it in this paper along with an analytic derivation taking higher-order terms into account than previous attempts from literature, in Sec.~\ref{sec:NW_non_monotonic}.

\subsection{\label{sec:NW_non_monotonic}Non-monotonical behavior of clustering coefficient}

\begin{figure}
\includegraphics[width=\columnwidth]{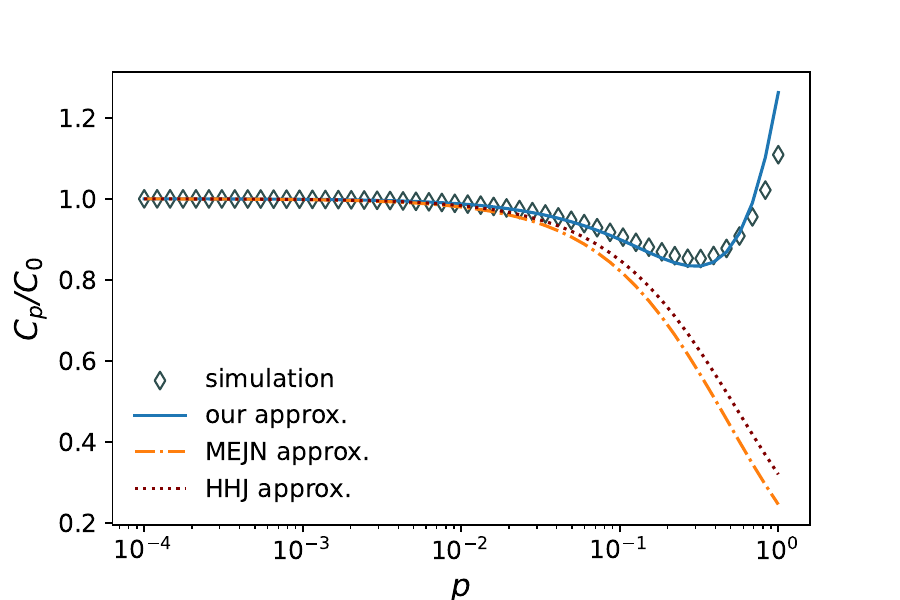}
\caption{Comparing the relative average clustering coefficient $C_p/C_0$ with respect to the 1D ring with $p=0$: our approximation in Eq.~\eqref{eq:C_approx}, Newman's approximation (MEJN)~\cite{NewmanBook1stEd}, Jo's approximation (HHJ)~\cite{HHJ_thesis}, and the simulation result (averaged over $500$ realizations) for the ONW model with $n=100$ and $k=40$.}
\label{fig:nonmonotonic}
\end{figure}

In Newman's derivation\footnote{Unfortunately, we have discovered that the part containing the formula and the derivation is omitted in the second edition of his book~\cite{NewmanBook2ndEd}.} of the average clustering coefficient for the NW model~\cite{NewmanBook1stEd}, neglecting appearance of new triangles generated by shortcuts is reasonable and justified in the large $n$ or sparse ($k_\mathrm{init} \ll n$) limit. A better approximation by considering an additional term for triangles generated by shortcuts is given by Jo~\cite{HHJ_thesis}, but it also neglects higher-order terms $\mathcal{O}(p^2)$ that will become more relevant as $p$ becomes larger. In this paper, we consider all of the possible triangles arising from the newly introduced shortcuts (including higher-order terms neglected in Ref.~\cite{HHJ_thesis}), and take the effect of disallowing multiple edges and self-loops into account. Our derivation predicts the average clustering coefficient of the ONW model with $n$ nodes and the initial degree $k_\mathrm{init}$ for a given value of $p$ as
\begin{widetext}
\begin{equation}
C = 3 \frac{\displaystyle \frac{nk_\mathrm{init}}{4} \left( \frac{k_\mathrm{init}}{2} - 1 \right) + \left( \frac{k_\mathrm{init}}{2} + 1 \right) \frac{k_\mathrm{init}}{4} \frac{k_\mathrm{init}p}{n-1-k_\mathrm{init}} n + \frac{k_\mathrm{init}^2p^2}{2}\frac{k_\mathrm{init}}{n}n + \frac{k_\mathrm{init}^2p^2}{2} \left(1 - \frac{k_\mathrm{init}}{n}\right)\frac{k_\mathrm{init}p}{n-1-k_\mathrm{init}}n}{\displaystyle \frac{1}{2}nk_\mathrm{init}(k_\mathrm{init}-1) + nk_\mathrm{init}^2p + \frac{1}{2}nk_\mathrm{init}^2p^2} \,.
\label{eq:C_approx}
\end{equation}
\end{widetext}

The first term in the numerator of Eq.~\eqref{eq:C_approx} corresponds to the initial number of triangles present in the underlying ring structure (when $p=0$), while the second, third, and fourth terms represent additional triangles introduced by one, two, and three shortcuts when $p>0$, respectively. We present more detailed derivation steps in Appendix~\ref{NW_CC_formula} for interested readers. This expanded consideration allows for a more accurate characterization of the system under investigation.
As demonstrated in Fig.~\ref{fig:nonmonotonic}, our formula clearly predicts the aforementioned \emph{increasing} part of the average clustering coefficient as $p$ increases by properly considering higher-order terms involving $\mathcal{O}(p^2)$, which is not captured by previous approximations~\cite{NewmanBook1stEd,HHJ_thesis}. Moreover, it is also possible to predict the ``turning point'' by differentiating the formula in Eq.~\eqref{eq:C_approx} with respect to $p$ and obtain the point where the derivative becomes zero by solving a cubic equation, although it is difficult to obtain a general closed-form solution as a function of $n$ and $k_\mathrm{init}$. The remaining discrepancy between the prediction of Eq.~\eqref{eq:C_approx} and the simulation result at $p \to 1$, shown in \ref{fig:nonmonotonic}, is likely caused by the approximation in our calculation detailed in Appendix~\ref{NW_CC_formula}.

\section{\label{sec:Harary_vs_NW}Equivalence between Harary and Newman-Watts models}

\begin{figure}
\includegraphics[width=\columnwidth]{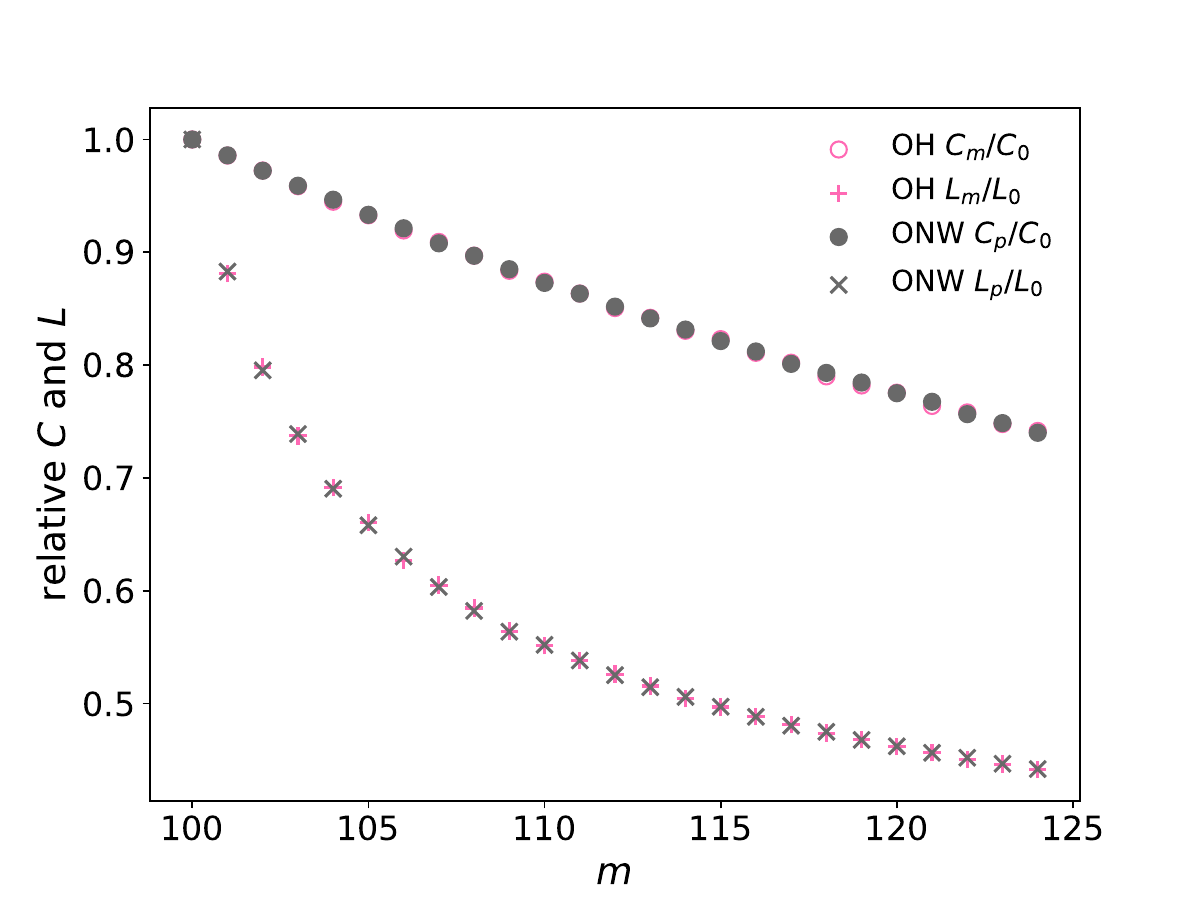}
\caption{The relative average clustering coefficient $C_m/C_0$ and $C_p/C_0$, and the relative average path length $L_m/L_0$ and $L_p/L_0$, both of which are normalized with respect to those ($C_0$ and $L_0$) in the original ring graph [$p=0$, or equivalently $m = nr/2$ according to Eq.~\eqref{eq:p_and_2m_over_nk}], are plotted against the total number $m$ of edges, in the cases of the OH ($r = 4$) and ONW ($k_\mathrm{init} = 4$) models composed of $n = 50$ nodes. All of the plots show the average values from $600$ independent realizations of each parameter value for a given network.}
\label{fig:Harary_vs_NW_CC_APL}
\end{figure}

Now we are ready to present that the OH and ONW models, developed in different eras (1960's versus 1990's) and mainly known in different disciplines (graph theory and computer science versus network science with physics background), are equivalent with a proper parameter mapping. First, note that both models have deterministic and stochastic parts; the former can involve regular 1D ring structures (corresponding to the cases of an even number $r$ for the OH model as introduced in Sec.~\ref{sec:originalHarary}), and the latter is composed of entirely random edge addition. Let us find the possibility of equivalence between the two models. 

Suppose we generate a network with $n$ nodes and $m$ edges. The initial 1D ring is characterized by an even number $r$ for the OH model and by $k_\mathrm{init}$ (usually an even number to avoid broken symmetry between the left and the right connections) for the ONW model. This initial $n$-ring exactly corresponds to the case of $r = 2m/n$ [the case~(\ref{case:2m_n_integer_r_even})] in the OH model and $p = 0$ in the ONW model. Therefore, a necessary condition for the equivalence between the two models is 
\begin{equation}
r = k_\mathrm{init} \,,
\label{eq:rk_equiv}
\end{equation}
where $r$ is the Harary parameter and $k_\mathrm{init}$ is the Newman-Watts parameter.

Then, let us examine the possibility of random edge addition for both models. For the OH model, the number of edges added randomly for the case~(\ref{case:2m_n_non_integer_2m_n_gtr_1_nr_even}) is $m - rn/2$, and for the ONW model, the number of edges added randomly is $p m_\mathrm{init} = npk_\mathrm{init} / 2$. As a result, by using the parameter relation in Eq.~\eqref{eq:rk_equiv},
the following parameter mapping completes the equivalence between the OH and ONW models: 
\begin{equation}
p = \frac{2m}{nr} - 1 \,,
\label{eq:p_and_2m_over_nk}
\end{equation}
where $p$ refers to the probability of random edge addition for the ONW model and $r$ is the connectivity parameter for the OH model. In summary, they are equivalent if the two conditions of Eqs.~\eqref{eq:rk_equiv} and \eqref{eq:p_and_2m_over_nk} hold. One important restriction is that $r$ should be an even number, but as we mentioned before $k_\mathrm{init}$ is usually restricted to an even number anyway in the ONW model for symmetry. An illustrative example is the OH model with $n=16$, $m=35$, and $r=4$ shown in Fig.~\ref{fig:Harary_examples}(d), which is equivalent to the ONW model with $n=16$, $k_\mathrm{int} = 4$, and $p=2m/(nr)-1 = 0.09375$.

One can check the equivalence of both models in Fig.~\ref{fig:Harary_vs_NW_CC_APL}, where the OH and ONW models show indistinguishable values for the average clustering coefficient and the average shortest path length, as long as the conditions in Eqs.~\eqref{eq:rk_equiv} and \eqref{eq:p_and_2m_over_nk} are met. On a technical note, the number of edges to be added in the ONW model is $m = (p-1)nr/2$ only on average for a given value of $p$, while the number of edges to be added in the OH model is exactly given as $m - rn/2$. The subtlety does not yield any notable difference in the case described in Fig.~\ref{fig:Harary_vs_NW_CC_APL}.

\section{\label{sec:conclusion}Conclusion and discussion}

This paper has revisited two network-generating models with different purposes and origins: the Harary graph~\cite{Harary1962} and the Newman-Watts model~\cite{Newman1999}. First, for both models, we have discovered the difference between the technical realizations described in the original literature and its modern implementation in the \texttt{NetworkX}~\cite{NetworkX} package. The differences might have been taken as negligible minor details, but throughout this paper, we have presented their impact on fundamental issues of stochasticity (in the Harary graph case) and on nontrivial structural differences represented by the degree distribution and the average clustering coefficient (in the Newman-Watts model case). 

During the investigation, we have also discovered some previously underemphasized facts about the Newman-Watts model, such as the non-monotonic behavior of the average clustering coefficient as a function of the edge addition probability supported by our analytic prediction. Finally, we have shown the equivalence between the Harary graph and the Newman-Watts model with adequate parameter mapping. The Harary graph was developed to maximize the connectivity, and the Newman-Watts model was developed to incorporate the small-world property. The fact that they are equivalent implies the closely interwoven relation between the two concepts: connectivity and small-world-ness, which has never been explicitly stated to our knowledge.  

Our study was initially inspired by Ref.~\cite{graph2nn}, which explores the relationship between graph structures in neural networks for machine learning and their prediction performance, including the (\texttt{NetworkX} version of) Harary graph as one of the examples. While the neural network based on the Watts-Strogatz small-world model~\cite{graph2nn,Xie2019} has gained wide recognition, the Newman-Watts model was also used in Ref.~\cite{Erkaymaz2017}, where the authors demonstrate the superior performance of the Newman-Watts small-world feedforward neural network in diabetes diagnosis compared to conventional feedforward neural networks (including the Watts-Strogatz one). They highlight the benefits of small-world network structures in enhancing diagnostic accuracy. 

To date, the application of the Harary model in artificial neural networks has not been explored in research, except for a case in Ref.~\cite{graph2nn} (again, we emphasize that the original version of the stochastic Harary graph can explore much wider parameter ranges, in contrast to the deterministic version used there). However, given the findings of our study regarding the small-world characteristics exhibited by the Harary model and its resemblance (including the equivalence for the even number of $r$) to the Newman-Watts model, there can be a potential for its application in the field of deep learning. Of course, beyond the application to deep learning, we hope to inform the existence of both models to wider scientific communities. Finally, we would like to emphasize the importance of paying attention to details in the generative mechanisms of model networks, and not overlooking possibilities of other properties of such models beyond their original intention, along with the relation to other models. 

\begin{appendix}
\section{\label{NW_CC_formula} Details of our formula for the average clustering coefficient of the Newman-Watts model}

To complete the equation and count all new triangles induced by shortcuts, we add three higher-order terms to the numerator of the approximation formula for the average clustering coefficient in Ref.~\cite{NewmanBook1stEd}, which is
\begin{equation}
C = 3\frac{\displaystyle \frac{nk_\mathrm{init}}{4} \left( \frac{k_\mathrm{init}}{2} - 1 \right)}{\displaystyle \frac{1}{2} nk_\mathrm{init} (k_\mathrm{init} -1) + nk_\mathrm{init}^2 p + \frac{1}{2}k_\mathrm{init}^2p^2} \,,
\label{eq:Newman_CC}
\end{equation}
where the possibility of new triangles generated by the shortcuts is ignored in the large sparse network limit (no modification in the numerator compared with the original number of triangles existing in the 1D ring with the degree $k_\mathrm{init}$, and only the denominator is modified by incorporating the triads formed by the shortcuts).
Without loss of generality, the average number of triangles per node should be equal to the number of such triangles that originate from any given node (in the probabilistic sense), which we refer to as node $v$.

Firstly, we count the triangles that can be formed by one shortcut. As described in Ref.~\cite{NewmanBook1stEd}, if we consider node $v$ and another node $w$ that is $d \in [k_\mathrm{init}/2+1,k_\mathrm{init}]$ steps apart on the circle, $(v,w)$ are connected by one or more paths of length two. Importantly, if this node pair $(u,w)$ becomes connected by a shortcut, those triads [from the path(s) of length two] are closed as triangles. For example, a shortcut between node $v$ and a node $k_\mathrm{init}/2+1$ steps away from $v$ would yield $k_\mathrm{init}/2$ new triangles, and a shortcut between $v$ and a node $k_\mathrm{init}/2+2$ steps away from it would yield $k_\mathrm{init}/2-1$ new triangles. Continuing this deduction up to the node $k_\mathrm{init}$ steps away, where a shortcut would yield a single new triangle, the total number of triangles potentially formed by one shortcut is $k_\mathrm{init}/2 + (k_\mathrm{init}/2-1) + (k_\mathrm{init}/2-2) + \cdots + 1$, resulting in $(k_\mathrm{init}/2+1)(k_\mathrm{init}/4)$ (the sum of an arithmetic sequence). This expectation value should then be multiplied by the probability that any particular pair of nodes is connected, $k_\mathrm{init}p/(n-1-k_\mathrm{init})$, and finally multiplied by the total number $n$ of nodes to obtain the expected number of triangles in the entire network. As a result, the total number of triangles formed by one shortcut is 
\begin{equation}
\left( \frac{k_\mathrm{init}}{2}+1 \right) \frac{k_\mathrm{init}}{4} \frac{k_\mathrm{init}p}{n-1-k_\mathrm{init}} n \,,
\label{eq:the_2nd_term}
\end{equation}
which comprises the the second term of the numerator of Eq.~\eqref{eq:C_approx}. Note that this deduction ignores the possibility of the triangle appearance from already existing shorcuts, which will be the subject of higher-order terms that will be discussed from now on.

Next, let us count the triangles that can be formed by two shortcuts. This occurs when a pair of shortcuts from node $v$ targets distant nodes denoted by $n_1$ and $n_2$, and there is already an edge connecting $n_1$ and $n_2$ as they happen to be close to each other (within the distance $k_\mathrm{init}/2$ on the ring, precisely), which results in the formation of a new triangle. The expected number of connected triads centered at node $v$ is given by
\begin{equation}
{k_\mathrm{init}p \choose 2} \approx \frac{k_\mathrm{init}^2 p^2}{2} \,, 
\label{eq:connected_triads}
\end{equation}
as the average number of shortcuts connected to each node is $k_\mathrm{init}p$. This formula in terms of the expected number of connected triads makes sense only for $p \gg 1/k_\mathrm{init}$, but at the other limit $p \ll 1/k_\mathrm{init}$, we can interpret the RHS of Eq.~\eqref{eq:connected_triads} as the \emph{probability} of a node to have exactly two shortcuts, given by 
\begin{equation}
{k_\mathrm{init} \choose 2} p^2 (1-p)^{k_\mathrm{init}-2} \approx  \frac{k_\mathrm{init}^2 p^2}{2} \,,
\label{eq:connected_triads_prob}
\end{equation} 
for small $p$. Lastly, the probability for the two nodes $n_1$ and $n_2$, connected to node $v$ via the shortcut, to be already connected in the original 1D ring or within the distance $k_\mathrm{init}/2$ is $k_\mathrm{init}/n$ (fix $n_1$ without loss of generality, and consider the location of $n_2$). Multiplying the probabilities and the total number $n$ of nodes, we obtain the third term of the numerator of Eq.~\eqref{eq:C_approx},
\begin{equation}
\frac{k_\mathrm{init}^2p^2}{2}\frac{k_\mathrm{init}}{n}n \,.
\label{eq:the_3rd_term}
\end{equation}

Finally, we count the triangles that can be formed by three shortcuts. This can be obtained by multiplying the expected number of connected triads formed by $(v,n_1)$ and $(v,n_2)$, which is $k_\mathrm{init}^2 p^2 / 2$ as discussed in the previous paragraph, and the probability of the distance between $n_1$ and $n_2$ being greater than $k_\mathrm{init}/2$ in the original 1D ring: $(1 - k_\mathrm{init}/n)$, and the probability of $(n_1,n_2)$ being connected by the third shortcut even they are distant to each other: $k_\mathrm{init}p/(n-1-k_\mathrm{init})$ (the term introduced in the one-shortcut case). Multiplying the probabilities and the total number $n$ of nodes, we obtain the fourth term of the numerator of Eq.~\eqref{eq:C_approx},
\begin{equation}
\frac{k_\mathrm{init}^2p^2}{2} \left(1 - \frac{k_\mathrm{init}}{n}\right)\frac{k_\mathrm{init}p}{n-1-k_\mathrm{init}}n \,.
\label{eq:the_4th_term}
\end{equation}

\end{appendix}

\begin{acknowledgments}
The authors thank Byunghwee Lee for discussing the specific realizations of the Newman-Watts model. This paper was written due to project-based learning (PBL) courses under the guidance of Jaemin Kong in Department of Physics, Gyeongsang National University (2022--2023). This work was supported by the National Research Foundation (NRF) of Korea Grants Nos. NRF-2021R1C1C1004132 (E.J.C. and S.H.L.) and NRF-2022R1A4A1030660 (S.H.L.).
\end{acknowledgments}

\end{document}